\documentclass[runningheads]{llncs}
\usepackage[T1]{fontenc}
\usepackage{graphicx}%
\usepackage{multirow}%
\usepackage{xcolor}%
\usepackage{textcomp}%
\usepackage{manyfoot}%
\usepackage{booktabs}%
\usepackage{algorithm}%
\usepackage{algorithmicx}%
\usepackage{algpseudocode}%
\usepackage{listings}%

\usepackage{draftwatermark}%
\SetWatermarkText{PREPRINT}%
\SetWatermarkScale{3}%
\SetWatermarkLightness{0.93}%

\usepackage{float}%
\usepackage{paralist}%
\usepackage{hyperref}%

\lstdefinestyle{cpp}{
  language=C++,
  basicstyle=\scriptsize\ttfamily,
  keywordstyle=\color{blue!80!black},
  stringstyle=\color{red!80!black},
  commentstyle=\color{green!50!black},
  numbers=left,
  numberstyle=\tiny\color{gray},
  stepnumber=1,
  upquote=true,
  keepspaces=true,
  flexiblecolumns=false,
  numbersep=5pt,
  showstringspaces=false,
  frame=single,
  tabsize=2,
  escapeinside={*@}{@*},
  literate=
    {__0}{{{\color{purple}0}}}1
    {__1}{{{\color{purple}1}}}1
    {__2}{{{\color{purple}2}}}1
    {__3}{{{\color{purple}3}}}1
    {__4}{{{\color{purple}4}}}1
    {__5}{{{\color{purple}5}}}1
    {__6}{{{\color{purple}6}}}1
    {__7}{{{\color{purple}7}}}1
    {__8}{{{\color{purple}8}}}1
    {__9}{{{\color{purple}9}}}1
    {::}{{{\color{gray}::}}}2
    {->}{{{\color{gray}->}}}2
}

\lstdefinestyle{exo}{
  language=Python,
  basicstyle=\scriptsize\ttfamily,
  keywordstyle=\color{blue!80!black},
  stringstyle=\color{red!80!black},
  numbers=left,
  numberstyle=\tiny\color{gray},
  stepnumber=1,
  upquote=true,
  keepspaces=true,
  flexiblecolumns=false,
  numbersep=5pt,
  showstringspaces=false,
  frame=single,
  tabsize=2,
  escapeinside={*@}{@*},
  literate=
    {__0}{{{\color{purple}0}}}1
    {__1}{{{\color{purple}1}}}1
    {__2}{{{\color{purple}2}}}1
    {__3}{{{\color{purple}3}}}1
    {__4}{{{\color{purple}4}}}1
    {__5}{{{\color{purple}5}}}1
    {__6}{{{\color{purple}6}}}1
    {__7}{{{\color{purple}7}}}1
    {__8}{{{\color{purple}8}}}1
    {__9}{{{\color{purple}9}}}1
}

\usepackage{color}

\begin{document}
\title{Effect of Abstractions and Prompting Strategies on LLM-Guided High-Performance Optimizations}
\titlerunning{Abstractions and Prompting Strategies in LLM-Guided Optimizations}
%
%
\author{Ji\v{r}\'i Klepl\orcidID{0000-0002-2231-4073} \and
Maty\'a\v{s} Brabec\orcidID{0009-0008-4470-2748} \and
Martin Kruli\v{s}\orcidID{0000-0002-0985-8949}}
\authorrunning{J. Klepl et al.}
%
\institute{Charles University, Prague, Czech Republic \\
\email{\{klepl,brabec,krulis\}@d3s.mff.cuni.cz}}
\maketitle              
%
\begin{abstract}
Code performance optimization is a vital aspect of modern software development, as it enables faster response times and reduced resource usage. These optimizations require a deep understanding of low-level hardware details and the intricacies of parallel processing, making them challenging even for experienced developers. With the advent of Large Language Models (LLMs), which are increasingly capable of generating and understanding code, there is growing interest in incorporating these models into automated code optimization processes. Traditionally, this automation involves transcribing the source code into a domain-specific representation that can be auto-tuned using grid search or machine learning algorithms, while adhering to strict rules and a limited set of feasible transformations to ensure verifiability. LLMs incorporate high-level code semantics and can thus perform transformations that go beyond verifiable automated optimizations. This paper investigates whether the traditional abstractions used in automated code optimization improve the performance and correctness of LLM-guided optimizations of parallel HPC applications. We evaluate this using the PolyBench benchmark suite and demonstrate that, in our evaluated setting, LLMs provided with specific optimization goals achieve better measured performance and validity rates when generating C code compared to creating computation pipelines and optimization schedules with established frameworks, suggesting that future development should explore alternative approaches for verifiable LLM-guided code optimization.
\end{abstract}

\section{Introduction}\label{sec:introduction}

Code performance optimization is an essential part of software development, ensuring quick response times and efficient use of computational resources. This challenge is particularly pronounced in high-performance computing (HPC) applications that often involve complex computations over large datasets. This challenge is addressed at multiple levels, from low-level optimizations performed by compilers like GCC and LLVM and tools based on polyhedral models~\cite{bondhugula2008pluto,grosser2012polly} to algorithm parallelization (typically performed by human experts) or auto-tuning frameworks, such as OpenTuner~\cite{ansel2014opentuner}. However, achieving optimal performance requires a deep understanding of the underlying hardware (e.g., memory hierarchies and parallelism ranging from SIMD to distributed computing) and the specific characteristics of the target application, such as exploitable invariants or optimizable memory access patterns.

Common optimization techniques that target memory hierarchies and parallelism include loop transformations such as tiling, loop interchange or skewing, unrolling, fusion, and parallelization strategies (e.g., OpenMP directives or SIMD vectorization). They are typically applied by the aforementioned polyhedral model-based frameworks or human experts, and then fine-tuned using auto-tuning frameworks. However, without deep expertise in hardware architectures and the mathematical models underlying these optimizations, applying these methods often leads to suboptimal performance~\cite{abdulla2012manual}.

To address this, domain-specific languages (DSLs) and frameworks, such as Halide~\cite{ragan2013halide}, Exo~\cite{ikarashi2022exocompilation}, and Noarr~\cite{klepl2024abstractions}, have been developed to simplify the process of specifying and tuning high-performance code for human programmers, replacing mathematical affine models with higher-level abstractions that better capture the semantics of computations and the intent of the optimizations.

In practice, applying these methods still requires human expertise to prepare optimization strategies that can be adapted to specific hardware and application characteristics. Alternative approaches based on polyhedral model optimizations are limited by the strict mathematical rules they must follow to ensure verifiability of the transformations and by speculative cost models.

Large language models (LLMs) have recently shown great promise in understanding and generating code, with models such as Codex~\cite{chen2021evaluating} demonstrating impressive results. Most of these works focus on web development or general programming tasks, often in languages that are not typically used for performance-critical applications. LLMs have also shown potential in generating HPC-based code~\cite{nichols2024performance,valero2023comparing} and optimizing existing code for better performance, usually by proposing directives to lead compiler optimizer or enable parallelization that are easily applicable and verifiable by a formal toolchain~\cite{chen2024ompgpt,chen2023lm4hpc}.
However, these approaches have limited expressiveness and may not achieve the optimal performance on tasks that require more complex or non-trivial optimizations, especially if the optimizations involve changes to the algorithm logic.

Recent studies have explored the use of LLMs to generate or optimize high-performance code using specialized prompting~\cite{brabec2025tutoring,chen2023vscuda,godoy2024large,palkowski2024gpt}, but these works typically focus on specific languages or frameworks.

Our work aims to explore the capabilities of LLMs across multiple domain-specific languages for high-performance parallel code development and optimization, as well as across different prompting strategies for generating the optimized code. The goal is to determine whether LLMs can effectively leverage the abstractions provided by these frameworks, or whether generating optimized parallel C code directly yields better performance and correctness.

As the representative LLM for our study, we chose GPT-5.1~\cite{openai2025gpt51}, one of the state-of-the-art models for code generation at the time of writing, offering reasonable performance while being priced favorably for our budget.

We formulate our objectives in the following research questions (RQs):
\begin{itemize}[]
    \item[\textbf{RQ1}] Can state-of-the-art frameworks for loop optimizations improve (e.g., add verification) LLM-guided optimization over direct code generation?
    \item[\textbf{RQ2}] Do more detailed prompts improve the performance and correctness of LLM-guided loop optimizations?
    \item[\textbf{RQ3}] Does following an explicit optimization plan generated prior to code generation improve the performance and correctness of LLM-guided optimizations?
    \item[\textbf{RQ4}] How do the approaches discussed in the previous questions compare in terms of performance gain, concurrency, precision, and overall prompt costs?
    \item[\textbf{RQ5}] Can LLMs apply optimizations that are not easily expressible in existing DSLs and frameworks reliably without compromising the optimized code?
\end{itemize}

To study these research questions, we forgo the use of an iterative process in which the LLM refines its outputs based on feedback from the compiler or runtime results. While such an approach is effective in practice~\cite{chen2023teaching,merouani2025agentic,shinn2023reflexion}, it also introduces additional variables that can obscure the studied factors. Furthermore, the iterative approach is often costly, and we aim to target reasonably priced solutions. Similarly, we consider a workflow that uses the LLM to both introduce a specific optimization framework (or DSL) and generate the optimized code, rather than one based on manual implementation in the target framework.

Answering the presented questions should help establish guidelines for using LLMs in high-performance parallel code optimization and clarify where existing optimization abstractions help or hinder LLM-guided optimization. In addition, we provide a replication package~\cite{replicationPackage} with the entire codebase, prompts, and experiment results to facilitate further research in this area.

The paper is organized as follows. Section~\ref{sec:background} outlines the common optimization techniques in the studied DSLs and frameworks.
Our proposed approach is formulated in Section~\ref{sec:methodology}.
Section~\ref{sec:evaluation} summarizes the experimental setup and results.
Section~\ref{sec:related-work} discusses related work, and Section~\ref{sec:conclusion} concludes the paper.

\section{Background}\label{sec:background}

Loop transformations and data layout transformations are common techniques, but they can be challenging to apply correctly and effectively. Let us demonstrate this on a well-known example --- the na\"{\i}ve matrix multiplication. Mathematically,
it can be defined as \(C' = C + A \cdot B\).
Where \(C\), \(A\), and \(B\) are matrices of sizes \(M\!\times\!N\), \(M\!\times\!K\), and \(K\!\times\!N\), respectively. An implementation in the C language may look like the following:

\begin{lstlisting}[style=cpp]
void matmul(int M, int N, int K, float *C, float *A, float *B) {
  for (int i = __0; i < *@\color{purple}M@*; i++)
    for (int j = __0; j < *@\color{purple}N@*; j++)
      for (int k = __0; k < *@\color{purple}K@*; k++)
        C[i**@\color{purple}N@* + j] += A[i**@\color{purple}K@* + k] * B[k**@\color{purple}N@* + j];
}
\end{lstlisting}

Applying loop tiling with tile sizes of 32 for both the \(i\) and \(j\) loops, changing the \(B\) matrix layout to be column-major for better memory access patterns, and unrolling the innermost \(k\) loop by a factor of 4 will have the following result:

\begin{lstlisting}[style=cpp]
for (int io = __0; io < *@\color{purple}M@*; io += __3__2)             // Tiled i loop
  for (int jo = __0; jo < *@\color{purple}N@*; jo += __3__2)           // Tiled j loop
    for (int i = io; i < min(io+32, *@\color{purple}M@*); i++)   // Inner i
      for (int j = jo; j < min(jo+32, *@\color{purple}N@*); j++) // Inner j
        for (int k = __0; k < *@\color{purple}K@*; k += __4) {       // Unrolled k loop
          C[i**@\color{purple}N@* + j] += A[i**@\color{purple}K@* + (k + __0)] * B[j**@\color{purple}K@* + (k + __0)];
          C[i**@\color{purple}N@* + j] += A[i**@\color{purple}K@* + (k + __1)] * B[j**@\color{purple}K@* + (k + __1)];
          C[i**@\color{purple}N@* + j] += A[i**@\color{purple}K@* + (k + __2)] * B[j**@\color{purple}K@* + (k + __2)];
          C[i**@\color{purple}N@* + j] += A[i**@\color{purple}K@* + (k + __3)] * B[j**@\color{purple}K@* + (k + __3)];
}
\end{lstlisting}

These optimizations prepare the code for possible parallelization of the outer loops or vectorization of the unrolled innermost loop. Although semantically equivalent, the code is significantly more complex compared to the original; thus, it is more error-prone and harder to maintain.
For this reason, various frameworks and domain-specific languages (DSLs) have been developed to express algorithms and their optimizations more easily and safely, without having to manually rewrite the whole code for each transformation, and possibly with automatically added vectorization or parallelization.

Examples of such frameworks that we selected as good representatives of different approaches are Halide~\cite{ragan2013halide}, Exo~\cite{ikarashi2022exocompilation}, and Noarr~\cite{klepl2024abstractions}. All three frameworks aim to facilitate the development of high-performance code through higher-level abstractions and optimizations, and offer ways to separate the definition of the algorithmic logic from the optimization and scheduling of computations. We illustrate the basic usage of each framework using the previously introduced na\"{\i}ve matrix multiplication example.

\subsection{Halide}\label{sec:background-halide}

Halide~\cite{ragan2013halide} is a domain-specific language (DSL) embedded in C++, primarily designed for image processing, stencil computations, and tensor computations. The following code snippet illustrates an example of how to implement and optimize matrix multiplication using Halide:

\begin{lstlisting}[style=cpp,morekeywords={Halide,Buffer,Func,Var,RDom,update,tile,parallel,fuse,vectorize,realize}]
Buffer<float> C(*@\color{purple}M@*, *@\color{purple}N@*), A(*@\color{purple}M@*, *@\color{purple}K@*), B(*@\color{purple}K@*, *@\color{purple}N@*);

// Algorithm definition
matmul(i, j) = C(i, j);
matmul(i, j) += A(i, k) * B(k, j);

// Schedule definition
matmul.update().tile(i, j, i_i, j_i, __3__2, __3__2).fuse(i, j, tile_idx)
               .parallel(tile_idx).vectorize(k, __4);
\end{lstlisting}

The core concept in Halide is the separation of the algorithm from its execution schedule. The algorithm is defined as a computation pipeline broken down into a series of \emph{functions} that describe how to compute the output values from the input values.

After defining a computation pipeline, Halide allows specifying various optimizations via a user-defined \emph{schedule} that describes how the computation should be executed. In the example above, we modify the second stage of the \lstinline{matmul} function by tiling the iteration space of \lstinline{i} and \lstinline{j} into $32\!\times\!32$, parallelizing two outer loops, and explicitly vectorizing the innermost loop. Even with these restrictions, the scheduling language is expressive enough to achieve significant performance improvements for a wide range of applications~\cite{mullapudi2016automatically,ragan2013halide}.

Since Halide is specifically designed for stencil-like computations and tensor pipelines, it is expected to perform well on such tasks when the computation and schedule are expressed idiomatically. However, the trade-off is that some algorithms are more challenging to express in an idiomatic Halide form, often requiring non-trivial workarounds that break the computation into iteratively applied steps. While such workarounds can help express algorithms with nested loops, they also increase the performance overhead, which has to be mitigated by Halide's optimizations and user-defined schedules.

\subsection{Exo}\label{sec:background-exo}

Exo~\cite{ikarashi2022exocompilation} is a Python-based framework for generating high-performance code for target platforms. Matrix multiplication can be implemented in Exo as follows:

\begin{lstlisting}[style=exo,morekeywords={@proc,size,@DRAM,seq}]
@proc
def gemm(
  *@\color{purple}ni@*: size, *@\color{purple}nj@*: size, *@\color{purple}nk@*: size,
  C: float[*@\color{purple}ni@*,*@\color{purple}nj@*]@DRAM, A: float[*@\color{purple}ni@*,*@\color{purple}nk@*]@DRAM, B: float[*@\color{purple}nk@*,*@\color{purple}nj@*]@DRAM
):
  for i in seq(__0, *@\color{purple}ni@*):
    for j in seq(__0, *@\color{purple}nj@*):
      for k in seq(__0, *@\color{purple}nk@*):
        C[i, j] += A[i, k] * B[k, j]
\end{lstlisting}

The core concept in Exo is defining computations as \emph{procedures} (\lstinline[style=exo]{@proc}) that describe the operations to be performed on the input data. The procedures can include standard control flow constructs such as (\lstinline[style=exo]{for}) loops and conditionals, allowing for a more general representation of algorithms compared to Halide; however, the language still strictly separates data computation from bounds and memory layout specifications --- for example, conditionals based on data values are not allowed. This ensures that the Exo engine can reason about the computation and apply optimizations safely, while sacrificing some flexibility in algorithm expression.

Similar to Halide scheduling, Exo allows specifying various optimizations through externally specified \emph{rewrite rules}~\cite{ikarashi2022exocompilation}. These rules are applied using pattern matching to specific operations or loop nests within the procedure, together with performing type inference, allowing the Exo engine to verify the correctness of the transformations before generating optimized C code. It also supports scheduling operations such as loop parallelization (using \lstinline[style=exo]{parallelize_loop}).

In contrast to Halide, Exo is designed for general loop-based computations. After performing the specified optimization rewrites and verifications, Exo uses the resulting structure to generate the optimized C code --- if no optimizations are applied, the expected generated C code for the above specification would result from simply replacing the specific Python syntax with C syntax.

The loop-based structure of Exo makes it more flexible than Halide, allowing it to express imperfectly nested loops, which is Halide's bane. However, with flexibility comes complexity. Thus, the optimizations available in Exo are more limited compared to Halide, unless the user provides additional information about the computation semantics to help Exo perform more complex transformations (such as assertions about loop bounds to enable tiling).

\subsection{Noarr}\label{sec:background-noarr}

Noarr~\cite{klepl2024abstractions} is a C++ header-only library that provides abstractions for defining data layouts (structures) and algorithms (structure traversal).
A Noarr implementation of matrix multiplication can be expressed as follows:

\begin{lstlisting}[style=cpp,morekeywords={noarr,bag,scalar,vector,traverser,V}]
auto C = bag(scalar<float>() ^ vector<'i'>(*@\color{purple}M@*) ^ vector<'j'>(*@\color{purple}N@*));
auto A = bag(scalar<float>() ^ vector<'i'>(*@\color{purple}M@*) ^ vector<'k'>(*@\color{purple}K@*));
auto B = bag(scalar<float>() ^ vector<'k'>(*@\color{purple}K@*) ^ vector<'j'>(*@\color{purple}N@*));

traverser(C, A, B) | [](auto idx) { C[idx] += A[idx] * B[idx]; };
\end{lstlisting}

Among the selected frameworks, Noarr uses the most lightweight and straightforward approach. It is a C++ library that provides template-based abstractions for defining and iterating data structures; thus, it allows full usage of C++ features. The core concept in Noarr is defining data structures using composable abstractions (like \lstinline[style=cpp,morekeywords={noarr,bag,scalar,vector,traverser,V}]{vector} or \lstinline[style=cpp,morekeywords={noarr,bag,scalar,vector,traverser,V}]{scalar}) that describe the layout and dimensions of the data, and performing computations using \emph{traversers} that iterate over the iteration space defined by the closure over their dimensions (sequential, OpenMP, TBB, and CUDA-based traversers are available at present). Noarr does not perform any semantic analysis of the computations, but imposes safety via type checking and dimension naming.

Since Noarr is a lightweight abstraction over C++, it is expected to behave similarly to hand-written C/C++ code for the same algorithm, given that the algorithm semantics are identical. Compared to Halide and Exo, Noarr is simpler in terms of setup and integration into existing C++ projects, but it lacks capabilities such as applying more complex transformations or performing verification. It also differs in that it abstracts indexation (e.g., \lstinline[style=cpp,morekeywords={noarr,bag,scalar,vector,traverser,V}]{C[i, j]} in Exo or Halide is represented as \lstinline[style=cpp,morekeywords={noarr,bag,scalar,vector,traverser,V}]{C[idx]} in Noarr, with no explicit indices). This, combined with dimensions of the structures and loop nests being named rather than positional, helps avoid certain classes of bugs related to incorrect indexing. It is thus expected that Noarr implementations will be less error-prone compared to analogs in Halide or Exo. However, since Noarr does not provide verification capabilities, logic errors in Noarr may be caught only at runtime rather than at compile time.

\subsection{PolyBench}\label{sec:background-polybench}

To test and evaluate our LLM-guided methods, we need to select and implement a set of coding problems that could benefit from HPC optimizations. The PolyBench~\cite{polybench} benchmark suite is a collection of 30 numerical computations commonly used in scientific computing and high-performance computing research. It includes a variety of algorithms from different domains, such as linear algebra, data mining, and stencil computations. All benchmarks are implemented in plain C, and the suite defines several problem sizes (\lstinline{MINI}, \lstinline{SMALL}, \lstinline{MEDIUM}, \lstinline{LARGE}, and \lstinline{EXTRALARGE}). In total, there are 150 benchmark variants for a given data type (e.g., \lstinline{float} or \lstinline{double}). The PolyBench suite is widely used for evaluating the performance of compilers, optimization techniques, and hardware architectures~\cite{ashouri2016cobayn,bondhugula2008pluto,grosser2012polly,merouani2025agentic,nichols2024performance,verdoolaege2013polyhedral}, so we consider it a suitable choice for our experiments.

\section{Methodology}\label{sec:methodology}

In our experiments, we instruct the LLM to automatically optimize the performance of the given problem while preserving the original semantics. Doing so in pure C code is challenging even for human programmers. As illustrated in Section~\ref{sec:background}, even a relatively simple optimization, such as performing tiling across the input dimensions of a matrix multiplication, changing the layout of the involved arrays, and applying unrolling and vectorization, can considerably increase code complexity.

The PolyBench optimization starts from existing C implementations. For framework-based optimization, our workflow begins by asking the LLM to express the same computation in the target framework or DSL before applying optimization prompts. The LLM produces multiple candidate representations, and we algorithmically select the most suitable one (further described in Section~\ref{sec:translation}). Automating this step also avoids introducing human-written optimizations that could bias the subsequent optimization stage.

Each code representation is subjected to two types of LLM-guided optimizations: \emph{direct} code optimizations and optimizations based on an \emph{abstract optimization plan}. In the direct optimizations, we try multiple prompts with different levels of detail. The abstract optimization plan approach is divided into two subsequent prompts: the first is to generate an abstract plan, and the second is to apply (follow) the generated plan to produce the optimized code. The idea is to help the LLM to better structure its thoughts before applying them to the code~\cite{kojima2022large}.
Furthermore, the abstract plan may be more readable to users, allowing for a better understanding of the optimizations applied by the LLM (mitigating the \emph{comprehension debt}~\cite{zhang2025beyond}).

\begin{figure}[htbp]
  \centering
  \includegraphics[width=.95\textwidth]{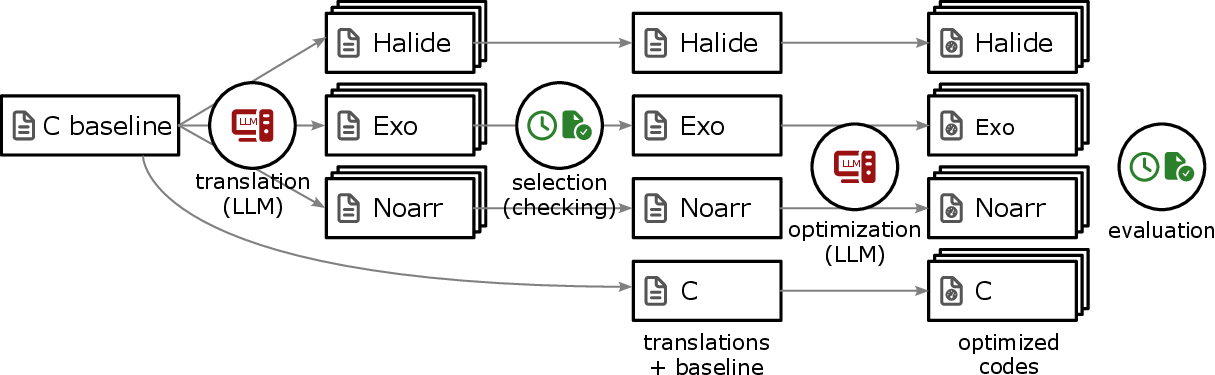}%
  \caption{Methodology overview}%
  \label{fig:methodology}
\end{figure}

A complete diagram of our methodology is shown in Figure~\ref{fig:methodology}. We address the individual parts in the following sections, while the technical details (including the prompts) are provided in the replication package~\cite{replicationPackage}.

\subsection{Translation}\label{sec:translation}

The LLM is requested to translate the original C code (a benchmark code from the PolyBench suite) into the target framework (or DSL), while preserving the original behavior as closely as possible. This stage produces the framework representation that the rest of the workflow will optimize. Ideally, the generated code follows the same structure as the original and performs the same computations in the same order (i.e., having the same memory access patterns), but using the constructs provided by the target framework.

In each request, the LLM is given (in the system message) the programming guide of the target framework, which is short enough to fit within the LLM's context window and specific enough to give the LLM a good idea of how to use the framework. Since the guides do not include complete reference documentation, we also provide the LLM with an example of translating the \texttt{GEMM} benchmark from PolyBench into the target framework, along with a list of framework-specific rules that the LLM should follow when generating the code. The goal of these additional resources is to explain further how to use the target framework correctly and to help the LLM avoid common pitfalls. We tailor the rules to each framework individually, based on our prior experience, and evaluate their effectiveness and sufficiency on a small subset of problems before proceeding to the full evaluation. We have used a representative set of four problems to tune the translation prompts: \texttt{GEMM}, \texttt{2MM}, \texttt{Floyd-Warshall}, and \texttt{Heat-3D}; these prompt-development benchmarks are excluded from the reported evaluation results.

We make five independent translation requests for each framework and benchmark combination to account for the variability in the LLM's outputs. Each translation candidate is first validated as described in Section~\ref{sec:validation}. Among the validated candidates, we select the one whose measured performance most closely matches the original C implementation across the \lstinline{SMALL}, \lstinline{MEDIUM}, and \lstinline{LARGE} input sizes from PolyBench. Only this selected translation is used as the input to the subsequent framework-based optimization experiments, so the reported framework results include both the quality of this translation step and the LLM's ability to optimize the selected representation.

\subsection{Direct optimization}\label{sec:optimization}

The direct optimization approach is similar to the translation step. The LLM is given the current representation produced by the workflow: the original C implementation for direct C optimization, or the selected translated representation for framework-based optimization. The prompt also includes the programming guide and specific rules of the respective framework when applicable. We make five independent optimization requests for each combination of benchmark, framework, and direct prompting strategy (including the C code).

The running system is briefly described in general terms as a modern x86-64 machine with a given number of CPU cores, along with the compiler version, the optimization flags, and the target framework specification. The LLM is then requested to optimize the code for performance while preserving its semantics. The prompts are tuned on the same representative set of four benchmarks as the translation prompts, as described in Section~\ref{sec:translation}. The optimization goals are specified by prompting the LLM to improve data locality, computation efficiency, and memory access patterns, while preserving the original code semantics and the boilerplate code used to set up the benchmarks and measuring their execution time.

For direct optimization, we use three prompting strategies in our endeavor to answer RQ2:
\begin{itemize}
  \item[\textbf{Na\"{\i}ve approach}] The LLM is simply prompted to optimize the provided code for performance while preserving its semantics without additional guidance.
  \item[\textbf{All-hints approach}] The LLM is provided with specific optimization hints (detailed below) to guide the optimization process.
  \item[\textbf{Choose-hints approach}] Is based on the \texttt{all-hints} approach, but the LLM is given an additional instruction to choose the most suitable hints from the provided ones and then apply them to the code.
\end{itemize}

The optimization hints provided to the LLM are divided into four categories. During prompt development, we also tried these categories as separate exploratory pilot prompts, but these pilot variants are not part of the reported design. In the reported experiments, the categories appear only as the combined hint set used by the \texttt{all-hints}, \texttt{choose-hints}, and \texttt{from-plan} approaches.

\begin{itemize}
  \item[\textbf{Cache}] Directs the LLM to apply tiling to the loops and reorganize the data layouts of the involved arrays to improve cache locality.
  \item[\textbf{Structure}] Directs the LLM to define temporary buffers of limited size (up to 50\% of the original arrays) to store intermediate results.
  \item[\textbf{Arithmetic}] Directs the LLM to consider applying algebraic transformations to reduce the number of arithmetic operations performed in the code. Also suggests using accumulator variables to minimize memory accesses within loops.
  \item[\textbf{Parallelism}] Directs the LLM to identify parallelization opportunities in the code and use appropriate constructs from the target DSL (or framework) to parallelize the computations across multiple CPU cores.
\end{itemize}

Each of these hints further asserts that the LLM should ensure that the optimizations preserve the original code semantics.

Together with the \texttt{from-plan} approach described below, the reported optimization design therefore consists of four prompting approaches crossed with four code representations (C, Exo, Halide, and Noarr).

\subsection{Abstract optimization plan}\label{sec:abstract-optimization-plan}

To address RQ3, we also investigate an alternative approach in which the LLM is first prompted (multiple times) to describe an abstract optimization plan without modifying the code, and then, in a second request, to choose from multiple optimization plans and apply the selected one to the original code. This approach should force the LLM to externalize a plan and reason about the optimizations before applying them to the code, potentially leading to better-structured and more effective optimizations~\cite{merouani2025agentic}.

\begin{figure}[htbp]
  \centering
  \includegraphics[width=0.8\textwidth]{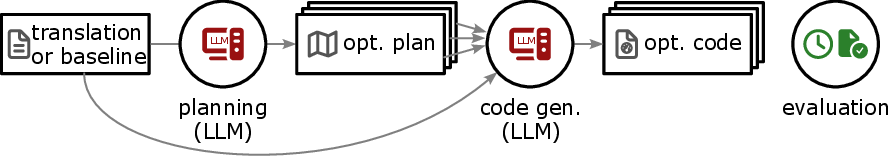}%
  \caption{Abstract plan optimization overview}%
  \label{fig:abs-plan}
\end{figure}

Figure~\ref{fig:abs-plan} illustrates the abstract optimization plan approach. The first request is based on the direct \texttt{all-hints} approach (the same four optimization hints are provided), but the model is prompted to describe an optimization plan. It differs in that the request contains the framework programming guide without the added framework-specific rules, and the LLM is instructed to describe the optimizations in abstract terms for later application. Similar to the translation step, the prompt includes an example of an abstract optimization plan for the \texttt{GEMM} benchmark with the expected format and level of detail.

The optimization plan is based on Halide and Exo principles --- namely, separating the data structures, computation pipelines, and optimization schedules. The LLM is expected to describe (in separate sections) the index domain (dimensions), the involved data structures and their layouts, any further algorithm parameters, and the computational steps to be performed (via loop nests representing quantification, and conditional blocks), and any relevant data dependencies that might affect parallelization or vectorization. Then, it is expected to describe the optimizations to be applied to the computation in terms of transformations of the described dimensions, data structures, and computational steps, via descriptive directives similar to those in the studied DSLs (such as tiling, loop interchange, unrolling, vectorization, and parallelization). The LLM is allowed to define its own directives as needed to express new optimizations, which aligns with the extensibility of Exo and Noarr and the intent to allow the LLM to leverage its full capabilities.

The system message for the second request is based on the na\"{\i}ve direct optimization prompts, with additional instructions to select the best plan from the previous request and apply it to the current code representation, instead of giving general optimization instructions. Similarly to the first request, the expected output is showcased on the \texttt{GEMM} benchmark in the prompt (using the target framework). The LLM is then provided with a user message containing the current code representation and all the previously generated plans.

\subsection{Validation}\label{sec:validation}

We validate the generated code by comparing its output with the output of the original unoptimized C code for the PolyBench-provided input sizes. We consider the generated code to pass validation if its output matches the original code's output within a small numerical tolerance (absolute tolerance $10^{-2}$ and relative tolerance $2\!\times\!10^{-7}$, chosen to account for precision errors) for all tested input sizes. If the generated code fails to compile or run for any input size, or produces mismatching output for any input size, it is considered invalid for this evaluation, and the specific failure category is recorded. We also impose an $8$-minute execution time limit, which is well beyond the expected execution time of any benchmark.

While this validation approach does not guarantee full semantic equivalence suitable for production use, the risk of false positives is low for the purposes of this study, as the PolyBench benchmarks are intentionally designed with non-symmetric input data and dimension sizes non-divisible by common tiling factors, making it unlikely that semantically incorrect code would pass on all tested input sizes by chance.

\subsection{Threats to validity}\label{sec:threats-to-validity}

One of the most common threats to validity when dealing with LLM outputs is their inherent stochasticity. We mitigate this by making multiple independent requests and selecting the output that best matches the original code in the translation step.

A related threat is the LLM's sensitivity to prompt wording and structure, which we mitigate by aligning the structure of the prompts with the provider's prompting best practices~\cite{openai2024prompt} and by tuning the prompts on a representative set of benchmarks (Section~\ref{sec:translation}) before proceeding to the full evaluation. We put great effort into minimizing unintentional bias in the prompts by constructing them in such a way that they do not favor any particular approach or framework and provide comparable levels of guidance and information across all approaches. Furthermore, we used the Grammarly tool to standardize the language quality and style across all prompts.

Finally, the LLM's familiarity with the benchmarks and their optimizations, as well as with the studied DSLs, may also threaten the validity of the results. We mitigate this by including multiple DSLs differing in design and usage, and by appending their respective programming guides, along with examples and LLM-specific rules, to the prompts. Based on the experiments used to tune the prompts (Section~\ref{sec:translation}), we observed no positive correlation between LLM competence and the framework's popularity or age.

\section{Evaluation}\label{sec:evaluation}

The evaluation was performed on a dual-socket system with Intel Xeon Gold 6130 processors, each comprising 16 cores with 2-way hyper-threading ($64$ threads in total). All benchmarks use double precision and were compiled with GCC 15.2, using \texttt{-O3}, \texttt{-march=native}, and \texttt{-mtune=native} optimization flags. The Halide code was linked against Halide v21.0.0. The performance was measured using a wall-clock timer, like in the PolyBench benchmark suite~\cite{polybench}. In all cases, only the 26 benchmarks not used for prompt design are considered for evaluation (as explained in Section~\ref{sec:translation}).

For the LLM requests, we used GPT-5.1 snapshot \texttt{gpt-5.1-2025-11-13}, reasoning effort set to \texttt{high} and verbosity set to \texttt{medium}~\cite{openai2025gpt51}; request construction is described in Section~\ref{sec:methodology}.

For each generated implementation, we averaged the execution time over three measured runs after two warm-up runs, which kept the measurement variance low on our system.

Generated implementations are validated using the procedure described in Section~\ref{sec:validation}. Invalid implementations are excluded from performance aggregates.

\subsection{Translation results}\label{sec:translation-results}

\begin{figure*}[tb]
  \centering
  \includegraphics[width=\linewidth]{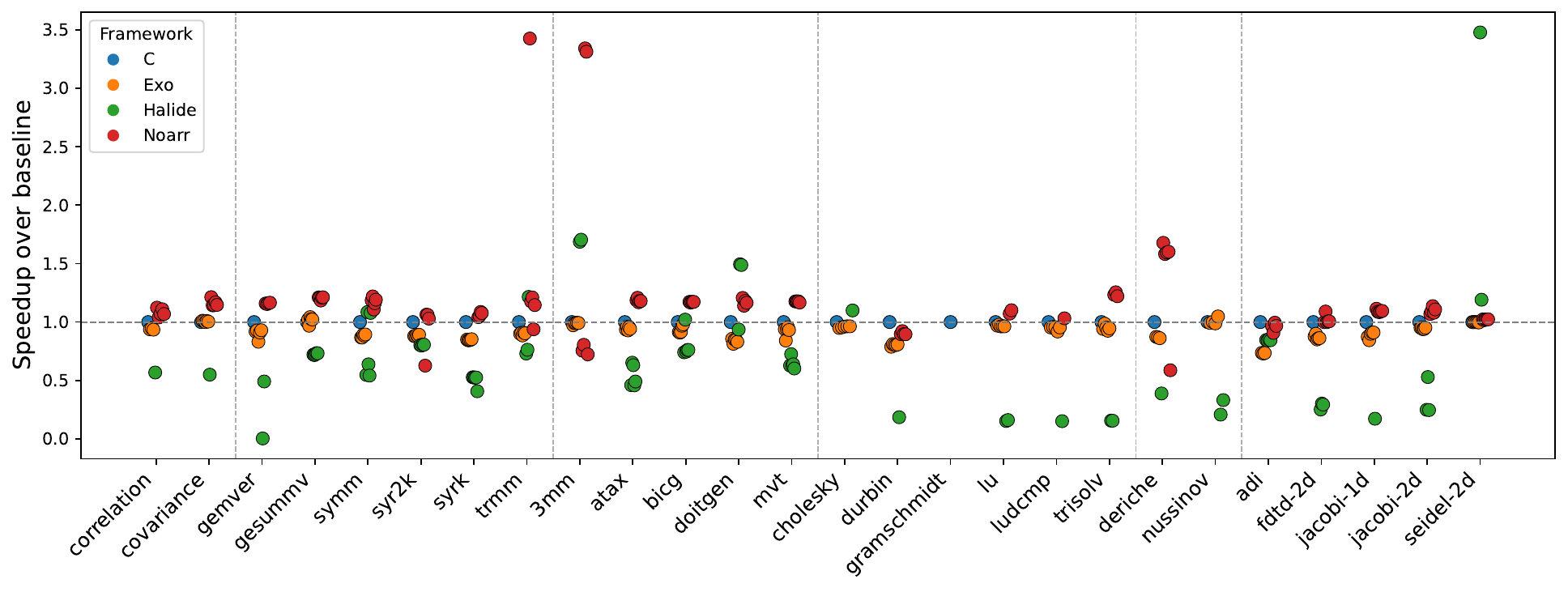}%
  \caption{Speedup of PolyBench benchmark translations to selected frameworks over the original C code. Each point represents one LLM translation of a given benchmark.}%
  \label{fig:translation-performance-scatter}
\end{figure*}

Figure~\ref{fig:translation-performance-scatter} compares the performance of the PolyBench benchmarks translated by LLM into Exo, Halide, and Noarr frameworks against the original C code without any optimizations applied. The $x$-axis represents the different benchmarks with horizontal lines separating each benchmark category as defined by the PolyBench suite. The $y$-axis shows the relative speedup of each translation over the baseline (emphasized by a line at $y=1$). The $y$ dimension of the displayed points shows the aggregated speedup over all input sizes defined by PolyBench for each benchmark, using the geometric mean, following standard practice~\cite{merouani2025agentic}. A lower distance from the baseline indicates better translation in terms of fidelity to the original benchmark.

The figure shows all translation attempts before selection. For the optimization experiments, each benchmark/framework pair contributes only the validated translation selected by the procedure in Section~\ref{sec:translation}: the candidate closest to C performance on the \lstinline{SMALL}, \lstinline{MEDIUM}, and \lstinline{LARGE} input sizes.

Most translations performing substantially worse than the C baseline belong to the Halide framework. In this workflow, this is mostly due to the LLM-generated Halide representations of cyclic data flows. Intuitive workarounds (e.g., running the Halide pipeline in a loop) often result in inefficient code that the LLM does not optimize until prompted to do so.

The higher variance in Noarr translations can be mostly attributed to Noarr defining array dimensions in reverse order compared to C\@; sometimes, the LLM prioritizes preserving the syntactic order of dimensions over semantic order, leading to accidental improvements or degradations in memory access patterns.

The LLM consistently fails to successfully translate the \lstinline{cholesky} and \lstinline{nussinov} benchmarks to Noarr and Halide due to more complex output formatting requirements, which the LLM fails to meet. The original benchmark outputs only a specific portion of the output array, which is overlooked by the LLM\@. For the \lstinline{gramschmidt} benchmark specifically, all LLM implementations fail due to numerical instability and arithmetic errors with all three frameworks.

The LLM achieves the highest success rate with Exo (which represents the most straightforward translation), failing only 12 times out of 130 attempts (including compile-time and runtime errors, and validation failures). Noarr fails in 30 and Halide in 61 attempts (of 130); however, a higher failure rate is expected for Halide in this workflow, as translating certain computations requires expressing them as a chain of function updates or numerous intermediate functions, which can easily lead to code that may pass validation, but is extremely inefficient. Overall, the results roughly correspond to the complexity of transcribing the source computations to each target framework.

\subsection{Optimization results}\label{sec:optimization-results}

\begin{figure}[htbp]
  \centering
  \includegraphics[width=\linewidth]{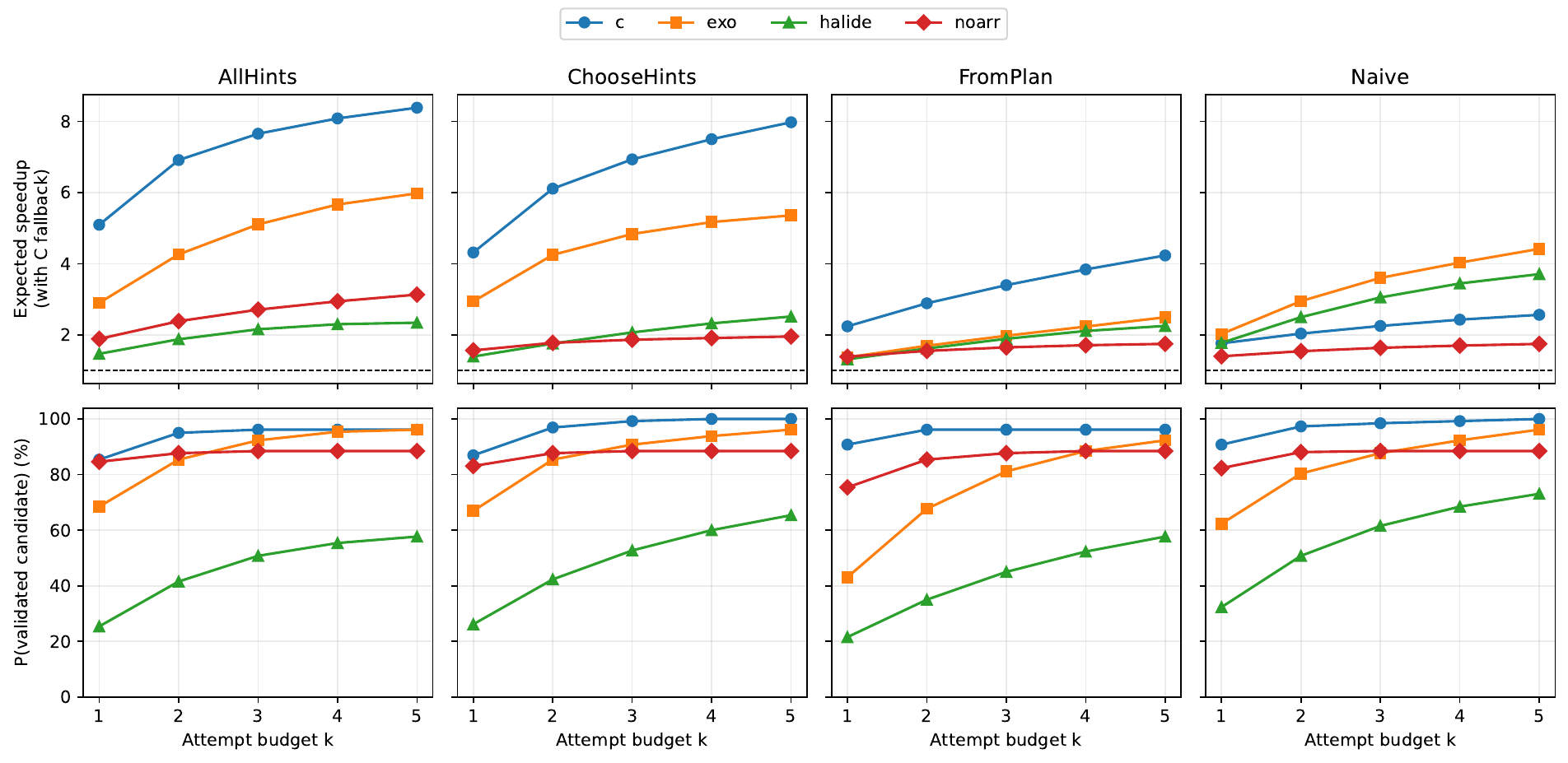}%
  \caption{Geometric mean of the best speedups and mean probability of at least one valid optimization attempt for a given prompting approach and framework over the C baseline for the \lstinline{EXTRALARGE} input sizes of the 26 benchmarks. The \(x\)-axis represents the budget \(k\) of independent attempts. Invalid attempts are counted in the validity denominator; speedups use the best valid candidate within the budget, with the C baseline available as a fallback.}%
  \label{fig:optimization-performance-best-of-k}
\end{figure}

The upper row of Figure~\ref{fig:optimization-performance-best-of-k} shows the geometric mean of the best speedup for a given prompting approach and framework pair for the \lstinline{EXTRALARGE} input size. The baseline is the corresponding unoptimized C implementation. The aggregation covers the 26 benchmarks not used for prompt design; for each benchmark, invalid generated implementations are ignored for speedup selection, and the C baseline is used as a fallback with speedup \(1.0\) if no valid implementation performs better.
The $x$-axis represents the budget \(k\) of independent attempts, while the $y$-axis shows the geometric mean of the best speedup achieved within that budget. The lower row shows the corresponding mean probability of obtaining at least one valid optimization within the same budget, so invalid attempts remain in the validity denominator.

We chose \lstinline{EXTRALARGE} as it models a case where performance typically matters the most. Plots for the other sizes and more detailed data (e.g., individual measurements and validation results) are available in our replication package~\cite{replicationPackage}.

The \texttt{all-hints} approach applied to direct C code generation performs the best overall performance, and the \texttt{choose-hints} approach performs only slightly worse in terms of performance (for Noarr, the performance degradation is more pronounced, suggesting optimization is less intuitive in this framework). We observed no indication that the LLM would pursue unsuitable optimization paths more frequently for \texttt{all-hints} compared to the \texttt{choose-hints} approach, despite the former being more aggressive in its optimization requests.

Contrary to initial expectations, the \texttt{from-plan} approach does not appear to provide clear performance benefits over the other two approaches, or even over the na\"{\i}ve approach. Moreover, the \texttt{from-plan} approach tends to actively hinder the performance compared to \texttt{all-hints} it is derived from. This suggests that the LLMs become more proficient in their own reasoning processes, and the users should not coerce them into deep reasoning artificially.

The validity rate is mostly independent of the prompting approach, with the exception of the \texttt{from-plan} approach, where the validity rate for the Exo framework is lower than for the other approaches. Direct C has the highest validity rate, independent of approach. For the Noarr framework, the LLM performs consistently slightly worse regardless of the given budget. At five attempts, the Exo framework becomes competitive with the C baseline in terms of validity rate, which suggests Exo code generation failures are more random than systematic in this setting (contrasting Noarr).

All prompting approaches show that the expected geometric mean speedup increases well with the number of independent attempts, with the exception of the Noarr framework. Disregarding Noarr, the budget of 5 independent attempts seems to be still in the steep part of the curve, suggesting little diminishing returns for the best-of-five approach. Achieving a better speedup is possible with more attempts at the cost of further diminishing returns.

Inspection of the outputs suggests that the LLM uses OpenMP parallelization for C and Noarr implementations and the built-in parallelization features of the Exo and Halide frameworks across all approaches. The observed tendency to apply parallelization is highest for the Halide framework, followed by C. This tendency is also influenced by the specificity of the optimization hints --- highest for the \texttt{all-hints} approach, followed by \texttt{choose-hints} and lowest for the \texttt{from-plan} approach. This suggests a higher cautiousness in the \texttt{from-plan} approach, which, however, is not reflected in the validity rates.

Across the 2,080 optimization attempts, 822 attempts were invalid. The failure-category tags record 454 compilation errors, 276 runtime errors, 91 numerical mismatches (invalid results), and 10 timeouts; these tags are diagnostic rather than disjoint, since one invalid attempt can fail differently across input sizes. Numerical mismatches and timeouts together contribute to only 5\% failure rate, making them a minor source of invalid optimizations.
Runtime errors were almost entirely concentrated in Halide (273 of 276), as it includes errors from its inner compilation pipeline. Compilation errors were the most common category overall. By framework, compilation-error tags were least frequent for C (33), followed by Noarr (65), Halide (167), and Exo (189).

\subsection{Comparison to prior work}\label{sec:comparison-to-prior-work}

We compare our best-performing approach (\texttt{all-hints} applied to direct C code generation) to prior work on LLM-guided optimizations, namely the approach of Merouani et al.~\cite{merouani2025agentic}, the most recent study with similar goals. They employ an iterative agentic framework to refine optimizations based on feedback from compiler and runtime measurements, and reported performance improvements of \(200\!\times\) or above on \lstinline{2MM}, \lstinline{3MM}, \lstinline{correlation}, \lstinline{covariance}, \lstinline{syr2k}, \lstinline{syrk}, and \lstinline{trmm}, with lower improvements on other PolyBench benchmarks at \lstinline{EXTRALARGE} input sizes. The work used a custom Tiramisu re-implementation for both the unoptimized baseline and the optimization target. Their best-of-five setting also uses 30 optimization iterations per run, for up to 150 explored schedules per benchmark instance.

Our results show that even with a non-iterative approach using only five LLM requests per benchmark and selecting the best optimization among them, it is possible to achieve results of a similar order of magnitude on several of these benchmarks. Against the original package-measured Tiramisu baseline, our best direct-C \texttt{all-hints} runs reach \(103.27\!\times\) on \lstinline{3MM}, \(47.42\!\times\) on \lstinline{correlation}, and \(73.11\!\times\) on \lstinline{covariance}. Furthermore, we observe high speedups on some benchmarks where their approach performed poorly, such as \lstinline{doitgen} (\(166.42\!\times\)) and \lstinline{symm} (\(53.68\!\times\)).

While their work demonstrates the potential of iterative refinement for LLM-guided optimizations, our findings indicate that a more direct approach (using pure C without iterative prompting) achieves strong performance in our setting.

We also observe much higher validity rates in our results compared to their work. While this could be attributed to differences in the strength of the LLM models used, we also note a success rate on Exo and Halide translations that is quite comparable to the validation rates observed by Merouani et al.~\cite{merouani2025agentic}, whose approach uses an abstraction related to Halide. This suggests that the incidence of invalid optimizations may also be attributed to the characteristics of the abstractions used.

\subsection{Answering research questions}\label{sec:research-questions-analysis}

\paragraph{\textbf{RQ1}: Do frameworks help?}
Section~\ref{sec:optimization-results} shows that using the frameworks in the context of our LLM-guided optimization workflow on the PolyBench suite does not provide clear performance or validity benefits over direct C code generation. It should be noted, however, that identifying an invalid optimization in a specifically designed framework can be easier than in C code due to the involved validation mechanisms, as shown by Merouani et al.~\cite{merouani2025agentic}. Conversely, these frameworks often impose strict rules that can limit the optimization opportunities the LLM can explore compared to direct C code generation, as we identified in Section~\ref{sec:comparison-to-prior-work}.

\paragraph{\textbf{RQ2}: Do specific hints help?}
Section~\ref{sec:optimization-results} indicates that providing more specific optimization goals to the LLM generally leads to better performance of the optimized code compared to na\"{\i}ve optimization requests. This trend is consistent across both framework-based and direct C code generation approaches.

\paragraph{\textbf{RQ3}: Does the abstract plan help?}
The goal of this research question was to determine if a two-step optimization process (first, generating an abstract optimization plan, then implementing it) improves optimization quality. We hypothesized that requesting the LLM to first propose multiple abstract strategies and subsequently select the most promising one to implement would reduce the likelihood of pursuing invalid or suboptimal paths.

Our results show that introducing the plan externalization phase does not generally improve performance compared to optimizing the code directly. Further investigation needed to fully analyze this effect is out of the scope of this work, but the recorded token accounting indicates that the LLM adapts its reasoning depth based on the given benchmark, which provides a more flexible approach than the tested plan externalization.

\paragraph{\textbf{RQ4}: Comparison of all approaches.}
From the results obtained for RQ1--RQ3, the specificity of optimization goals (RQ2) appears to have the greatest positive impact on the performance of LLM-guided loop optimizations. Specifically, the \texttt{all-hints} approach yields the best performance across all frameworks, with a minor exception for Halide.

Contemporary frameworks and DSLs do not clearly improve validity rates compared to direct C code generation in this workflow; however, the Exo framework comes close to matching the validity rate of direct C code generation after 5 independent attempts, regardless of the prompting approach.

\paragraph{\textbf{RQ5}: Are LLMs constrained by existing optimization frameworks?}
The results discussed in Section~\ref{sec:comparison-to-prior-work} indicate that LLMs can indeed reliably produce optimizations that are not easily expressible in optimization directives available in existing frameworks.
This suggests that future research should focus on developing abstractions better aligned with the capabilities of LLMs, achieving verifiability without relying on prefabricated optimization templates.

\section{Related Work}\label{sec:related-work}

Automated code optimization has historically relied on analytical frameworks, most notably the polyhedral model. Tools such as Pluto~\cite{bondhugula2008pluto} and Polly~\cite{grosser2012polly} leverage mathematical abstractions to perform aggressive loop reorganizations, including tiling, skewing, and interchange. While effective, these techniques are constrained by strict static analysis requirements, limiting their ability to apply more complex transformations.

For greater flexibility, domain-specific languages (DSLs) like Halide~\cite{ragan2013halide}, Exo~\cite{ikarashi2022exocompilation}, Noarr~\cite{klepl2024abstractions}, and Tiramisu~\cite{baghdadi2019tiramisu} decouple algorithms from execution schedules. This separation facilitates auto-tuning, where schedulers search for optimal configurations using beam search or learned cost models~\cite{adams2019learning,mullapudi2016automatically}. However, these approaches are often computationally demanding, especially when targeting diverse hardware backends.

The advent of Large Language Models (LLMs) has transformed code generation~\cite{jiang2024survey,roziere2023code}. Despite their success in general programming, empirical studies indicate that off-the-shelf models struggle to produce efficient parallel code for high-performance computing (HPC) tasks~\cite{godoy2024large,nichols2024can,valero2023comparing}. To bridge this gap, researchers have explored various enhancement strategies.

One direction is fine-tuning of specialized models. Examples include HPCCoder~\cite{nichols2024hpc}, which targets parallel code generation, and the LM4HPC framework~\cite{chen2024ompgpt,chen2023lm4hpc} for OpenMP pragmas. Beyond standard fine-tuning, reinforcement learning (RL) has been applied to discover novel algorithms (AlphaDev~\cite{mankowitz2023faster}) or to align models with performance feedback (RLPF~\cite{nichols2024performance}). However, these methods are resource-intensive, require access to proprietary weights, and risk obsolescence as foundation models evolve.

Another direction uses prompt engineering. Studies show that expert-crafted hints, source-to-source directives, and zero-shot reasoning strategies can significantly improve performance without retraining~\cite{brabec2025tutoring,chen2023vscuda,kojima2022large,palkowski2024gpt}. This approach relies on user expertise to formulate effective prompts.

To automate this process, agentic frameworks employ iterative feedback loops using compiler or runtime signals to refine outputs~\cite{chen2023teaching,shinn2023reflexion}. A prominent example is the work by Merouani et al.~\cite{merouani2025agentic}, which integrates LLMs with Tiramisu to generate verifiable schedules. While robust, such agentic approaches could be very costly, often requiring dozens of compilation cycles (and prompts) per task.

We leverage generalized optimization hints and generated optimization plans to guide the LLM, achieving competitive speedups via cost-efficient, non-iterative prompting. Compared to prior work, our approach focuses on investigating multiple DSLs and frameworks, as well as direct C code generation, and different prompting strategies to identify the most effective methods for LLM-guided code optimization.

\section{Conclusion}\label{sec:conclusion}

In this work, we present a study on using a large language model to optimize code for high-performance parallel applications in conjunction with frameworks that aim to ease the optimization process for both human programmers and automated tuning systems. In our evaluated setting, direct C generation produced stronger measured results than using these frameworks, both in terms of performance and correctness. We also observed that, for the evaluated workflow, the specificity of the prompt plays a larger role than whether the LLM is tasked to optimize code directly in C/C++ or through tools for user-guided optimization, such as Exo, Halide, or Noarr.

Two important implications arise. First, it suggests that prompt engineering is a critical factor in the success of LLM-guided code optimization. Second, it indicates that future research should focus on tailoring DSLs and alternative approaches to better align with LLM capabilities, potentially leading to new tools for code optimization that leverage the strengths of both approaches.

\begin{credits}
\subsubsection{\ackname} This paper was supported by Charles University institutional funding SVV (grant 260821) and by the Johannes Amos Comenius Programme (P JAC, Natural and anthropogenic georisks) project CZ.02.01.01/00/22\_008/0004605.

\subsubsection{\discintname}
The authors have no competing interests to declare that are
relevant to the content of this article.
\end{credits}
%
%
%
\bibliographystyle{splncs04}
\bibliography{bibliography}
\end{document}